# Singularity of Navier-Stokes Equations Leading to Turbulence

**Hua-Shu Dou**
Faculty of Mechanical Engineering and Automation,
Zhejiang Sci-Tech University, Hangzhou, Zhejiang 310018, China
Email: huashudou@yahoo.com

**Abstract:** Singularity of Navier-Stokes equations is uncovered for the first time which explains the mechanism of transition of a smooth laminar flow to turbulence. It is found that when an inflection point is formed on the velocity profile in pressure driven flows, velocity discontinuity occurs at this point. Meanwhile, pressure pulse is produced at the discontinuity due to conservation of the total mechanical energy. This discontinuity makes the Navier-Stokes equations be singular and causes the flow to become indefinite. The analytical results show that the singularity of the Navier-Stokes equations is the cause of turbulent transition and the inherent mechanism of sustenance of fully developed turbulence. Since the velocity is not differentiable at the singularity, there exist no smooth and physically reasonable solutions of Navier-Stokes equations at high Reynolds number (beyond laminar flow). The negative spike of velocity and the pulse of pressure due to discontinuity have obtained agreement with experiments and simulations in literature qualitatively.

## I Introduction

Turbulence is one of the most important scientific problems in physics. Reynolds pioneered the work for pipe flow in 1883, which proved that there are two types of flow states, laminar flow and turbulence [1]. Since then, substantial work has been done in theories, experiments and simulations on turbulence during the past 135 years or so. Although great progress has been made, the physical mechanism of turbulence is still poorly understood.

Over the years, it has been suggested that transition of laminar flow to turbulence is caused by the instability of laminar flow [2, 3]. The flow between two parallel plates is a classical flow problem. Heisenberg [2] obtained an approximate solution of the stability equation by the derivation of the linearized Navier-Stokes equations and gave the boundary of stability on the diagram of wave number versus the Re number. Lin [3] proved by mathematical asymptotic analysis that the flow between the two parallel plates would be unstable and obtained the critical Re number to be 8000. However, whether or not turbulence would occur after flow instability setting in was not clarified. Later, several researchers obtained similar results which approach the accurate value gradually. Orszag obtained the critical Re number to be 5772 by calculating with the spectral method [4]. This is the most accurate value recognized in the literature. However, the critical Re number of turbulent transition obtained from experiments is about 1000. This inconsistency between the theory and the experiment has been perplexing the understanding of turbulent transition [4-7].

A large quantity of experimental data and numerical calculations show that turbulence is a local phenomenon when it first starts [8-14]. With the increase of Re and the development of disturbance, inflection point first appears on the velocity profile in laminar flow, followed by the formation of hairpin vortices which finally leads to the turbulence spots. During the transition, the velocity profile is subjected to a continuous modification from a laminar profile to a turbulent profile [7,11,14-18]. It is found that the appearance of the velocity inflection point is a key step of turbulent transition [11-18] and recently experiment confirmed that turbulence is indeed sustained by an inflection point instability [13].

In the experiment of boundary layer flow [9], it was discovered that breakdown of the laminar flow starts with the appearance of *negative* spikes on the streamwise velocity temporal traces under influence of disturbances. The evolution is from one spike, then the spike doubled, tripled, etc. downstream. In some experiment and simulation, it was shown that the maximum amplitude of the negative spike reaches more than 50% of the streamwise velocity [19]. Kachanov reviewed the progress of researches on transition in boundary layer flow, and discussed the transition phenomenon and the roles of spikes in the flow breakdown [20]. However, the origin of the spike is still not fully understood up to present.

In the experiments and simulations, it is also found that negative spikes of streamwise velocity appear during the late stage of the transition of plane Poiseuille flow or channel flows [17, 21-24]. The simulation results showed that the appearance of the negative spike in streamwise velocity is accompanied by the distortion/inflection of velocity profile and the high-shear layer sloped to the wall [17, 18].

Dou and co-authors proposed an energy gradient theory for the study of flow stability and turbulent transition [25-34]. It is found that the inflection point on the velocity profile for the pressure driven flow is a singular point hidden in the Navier-Stokes equations. It is obtained for pressure driven flows that ***the necessary and sufficient condition for the turbulent transition is the existence of an inflection point on the velocity profile*** [25,27,29]. However, the characteristics of velocity change at the inflection point have not been given, and the discontinuity of flow parameters has not been accurately described.

Large quantities of theoretical, experimental and direct numerical simulation (DNS) results show that the Navier-Stokes equations are the governing equations to correctly describe both the laminar flow and the turbulent flow. As such, the physical mechanism of turbulent transition should exist in the Navier-Stokes equations. The mechanism of turbulence generation should be consistent and unique, no matter what type of turbulences such as wall bounded turbulence or free boundary turbulence.

However, whether the three-dimensional (3D) incompressible Navier-Stokes equations have unique smooth (continuously differentiable) solutions is still not known. And the physics hidden behind the equations attracted many mathematicians, physicists and engineers working on this problem for long time [35-36]. To solve this problem from mathematics and physics is very imperative

As is well known, there is discontinuity in the time-averaged Navier-Stokes equations during the transition from laminar flow to turbulence, as commonly observed from the drag coefficient [37-38]. On the other hand, it was found in the experiments that ***there is discontinuity of streamwise velocity*** in full developed turbulent flow [39-40]. However, this very important phenomenon of velocity discontinuity has not been obtained sufficient attention during the past thirty years. The mechanism of discontinuity of streamwise velocity is still not clear. Simultaneously, the origin of fluctuation of wall pressure is clear neither in turbulence. It is very important to determine the origin of discontinuity in the governing equations from mathematics and physics, to understand the mechanism of turbulent transition as well as the mechanism of fully developed turbulence.

In this paper, the singularity of Navier-Stokes equations is analyzed through the derivation of the Navier-Stokes equations and the analysis of the velocity profile for plane Poiseuille flow. Negative answer on existence and smoothness of the solution of the Navier-Stokes equations is given. Then, the proposed theory is compared with experiments and simulations in literature. Finally, the origin of turbulent transition and the mechanism of fully developed turbulence sustenance are discussed.

## 2 Discovering the Singularity of Navier-Stokes equations

### 2.1 Discontinuity Existing in the Transition of Laminar Flow to Turbulence

The time-averaged Navier-Stokes equations of incompressible fluid are as follows for laminar and turbulent flows, respectively [37-38],

$$\rho(\frac{\partial \overline{\mathrm{u}}}{\partial t}+\overline{\mathrm{u}}\cdot\nabla\overline{\mathrm{u}})=-\nabla\overline{p}+\mu\nabla^2\overline{\mathrm{u}}\,. \tag{1}$$

$$\rho(\frac{\partial \overline{\mathrm{u}}}{\partial t}+\overline{\mathrm{u}}\cdot\nabla\overline{\mathrm{u}})=-\nabla\overline{p}+\mu\nabla^2\overline{\mathrm{u}}+\nabla\cdot\overline{\tau}_{\mathrm{t}}\,. \tag{2}$$

where $\mu$ is the fluid dynamic viscosity, $\overline{p}$ is the static pressure, $\rho$ is the fluid density, $\overline{\mathrm{u}}$ is the velocity vector, and $\overline{\tau}_{\mathrm{t}}$ is the turbulent stress tensor. It is noted that the time-averaged Navier-Stokes equations of incompressible fluid for laminar flow is the same as that with not time-averaged equations.

Thus, it can be observed from Eqs.(1) and (2) that there is discontinuity from laminar flow to turbulence in the time-averaged Navier-Stokes equations due to existence of $\overline{\tau}_{\mathrm{t}}$. Therefore, there must be a discontinuity of flow physics in laminar flow from which the flow transits to turbulence (Fig.1). Figure 2 shows the typical velocity profiles for laminar, transitional and turbulent flows for plane Poiseuille flow. The discontinuity from Eq.(1) to Eq.(2) indicates that there is singular point in the time-averaged Navier-Stokes equations of the laminar flow, via which discontinuity occurs. ***If the mechanism of the said discontinuity is found, the physics of turbulent transition would be consequently understood.*** As is well known, the laminar flow is smooth. A laminar flow can only transit to turbulence at a higher Reynolds number and under certain large disturbance [25, 27]. At what condition, the discontinuity occurs, how it is produced, and what result it will generate, these are just what we want to know. Since the transition of laminar flow to turbulence is dominated by the Navier-Stokes equations, the physics of discontinuity must hide in the equations under some flow conditions which are suitable for the transition occurrence.

### 2.2 Axiom for the Gradient of the Total Mechanical Energy and Velocity

In order to analyze the discontinuity occurring in laminar flow from the viewpoint of energy variation, a preliminary work for proving an axiom is given first. In the following, the principle of energy conservation is employed to analyze the difference of the variation of the total mechanical energy between inviscid flow and viscous flow along a streamline. For incompressible flow, the total mechanical energy ($E$) is constant along a streamline ($x$) in inviscid flow, $\partial E/\partial x = 0$, while it will decrease along a streamline in viscous flow, $\partial E/\partial x < 0$, due to viscous friction. This statement is only correct for pressure driven flows, for example, plane Poiseuille flow and pipe Poiseuille flow [25, 27]. For shear driven flow, this statement is not correct due to the input of external work, for example, in plane Couette flow [26, 28].

As is well known, for viscous fluid flow, fluid particle consumes energy with the flow forward due to viscous friction. For incompressible fluid in laboratory system, fluid particle always flows from the position with higher total mechanical energy to the position with lower total mechanical energy, if there is no work input. Along a streamline direction (x expresses the streamline direction and $u$ expresses the velocity), if $\dfrac{\partial E}{\partial x} < 0$, $u > 0$; if $\dfrac{\partial E}{\partial x} > 0$, $u < 0$; if $\dfrac{\partial E}{\partial x} = 0$, $u = 0$. Therefore, we obtain the following axiom (1):

**Axiom (1): In laboratory system, if the gradient of the total mechanical energy along a streamline becomes zero, the flow velocity is zero.**

This axiom is correct only for pressure driven flows, and not for shear driven flows.

### 2.3 Exploring the Singularity Point of the Navier-Stokes Equations

Now, we pay our attention to a three-dimensional laminar flow between two parallel walls as shown in Fig.3 (plane Poiseuille flow). The width of two plates in spanwise direction is infinite. The height between the two plates is 2h. The wall is set as no-slip condition. The incoming flow is a laminar velocity profile. The exact solution of the velocity for laminar flow is a parabolic velocity distribution along the height for Newtonian fluid [37-38]. We put this smooth velocity distribution in the flow field as the initial condition. Then, we will observe the variation of the velocity distribution with time under external disturbances, as in simulations and experiments [11,15,17,18]. With the flow development from the interaction of the base flow with the disturbance exerted on the flow, the velocity profile can be modified depending on the Reynolds number and the disturbance. In the simulations and experiments, an inflectional velocity profile has been observed during the transition of laminar flow to turbulent flow, as shown in Fig.2 [11,15,17,18].

In the following, we will use the analysis of Navier-Stokes equations to explore the mechanism of the flow discontinuity from laminar flow to turbulence in plane Poiseuille flow.

For the pressure driven flow between two parallel plane walls (Fig.3), the continuity and the unsteady Navier-Stokes equations of the laminar flow for incompressible fluid can be written as follows [37-38],

$$\nabla \cdot \mathrm{u} = 0 . \tag{3a}$$

$$\rho(\frac{\partial \mathrm{u}}{\partial t} + \mathrm{u} \cdot \nabla \mathrm{u}) = -\nabla p + \mu \nabla^2 \mathrm{u} + \mathrm{f} . \tag{3b}$$

Here $\mathrm{u}$ is the velocity vector, $p$ is the static pressure, $\rho$ is the fluid density, $\mu$ is the dynamic viscosity and $\mathrm{f}$ is the gravitational force. The wall boundary condition is,

$$\mathrm{u} = 0 . \tag{3c}$$

With the identity

$$\mathrm{u} \cdot \nabla \mathrm{u} = \nabla(\frac{1}{2} \mathrm{V}^2) - \mathrm{u} \times (\nabla \times \mathrm{u}) , \tag{4}$$

Eq.(3b) is rewritten as

$$\rho \frac{\partial \mathrm{u}}{\partial t} + \nabla E = \mu \nabla^2 \mathrm{u} + \rho \mathrm{u} \times (\nabla \times \mathrm{u}) \,. \tag{5}$$

Here, $V$ is magnitude of the total velocity, $E = p + \frac{1}{2}\rho V^2 + G$ is the total mechanical energy of unit volumetric fluid, and $-\nabla G = \mathrm{f}$ is the gravitational force. In engineering practice, the gravitational force is often neglected.

For parallel flows, the Eq.(5) can be written as follow along the streamwise direction (x direction),

$$\rho \frac{\partial u}{\partial t} + \frac{\partial E}{\partial x} = \mu \left( \frac{\partial^2 u}{\partial x^2} + \frac{\partial^2 u}{\partial y^2} \right) \tag{6}$$

where $u$ is the velocity component in streamwise direction ($x$ direction).

For parallel flow, $\left|\frac{\partial^2 u}{\partial x^2}\right| << \left|\frac{\partial^2 u}{\partial y^2}\right|$, $\left|\frac{\partial^2 u}{\partial x^2}\right| \approx 0$, Eq.(6) is rewritten as,

$$\rho \frac{\partial u}{\partial t} + \frac{\partial E}{\partial x} = \mu \frac{\partial^2 u}{\partial y^2} \tag{7}$$

Omitting the term $\partial^2 u / \partial x^2$ from Eq.(6) to Eq.(7) will not affect the conclusions obtained in this study, for which further clarification and statement will be given in the following parts.

When there is an inflection point on the velocity profile, the flow status can be classified as the following three cases (Fig.4), (a) inflection point is not located at the wall, $u'' < 0$ at the wall; (b) inflection point is located at the wall, $u'' = 0$ at the wall; (c) inflection point is not located at the wall, but the velocity profile shows $u'' > 0$ at the wall. Here, $u''$ stands for the second derivative of the velocity to the direction normal to wall, $\partial^2 u / \partial y^2$.

In the following, the behavior of the inflection point is first analyzed for case (a), as shown in Fig.5. The incoming velocity profile is a smooth laminar flow, $u > 0$ and $u'' < 0$, and there is no inflection point on the velocity profile (Fig.5(a)). With the flow forward, as the interaction of the base flow with the disturbance progresses, the velocity profile is distorted. The magnitude of $u''$ in some place on the velocity profile, $|u''|$, is reduced. At some location such as at position $A$, $u''$ first becomes zero, and an inflection point appears at position $A$, $y = y_A$, $u_A'' = 0$ (Fig.5(b)). For plane Poiseuille flow, this position occurs generally at about 0.58 fraction of the width from centerline from theory and experiment [11, 25, 27].

In the following, we will discuss the solution of Eq.(7) for steady flow and unsteady flow conditions, respectively.

**(1)Steady flow.** For steady flow, Eq.(8) is obtained from Eq.(7) due to $\frac{\partial u}{\partial t}=0$,

$$\frac{\partial E}{\partial x}=\mu\frac{\partial^2 u}{\partial y^2} \tag{8}$$

For the given inflectional velocity profile in Fig.5 (b), $u_A''=0$ is obtained at the inflection point *A*. Then, we immediately obtain Eq.(9) from Eq.(8),

$$\frac{\partial E}{\partial x}=0 \quad \text{at} \quad y=y_A. \tag{9}$$

In previous section, we have discussed that ***the velocity in viscous flow is generated by the gradient of the total mechanical energy along the streamline.*** Since the gradient of the total mechanical energy along the streamline at point A is zero from Eq.(9), the velocity at the inflection point *A* should be zero in the laboratory system according to Axiom (1). Therefore, we obtain (for steady flow),

$$u_A=0. \tag{10}$$

**(2) Unsteady flow.** For the given inflectional velocity profile in Fig.5 (b), if this velocity profile is for an **unsteady flow** $\frac{\partial u}{\partial t}\neq 0$, the Eq.(7) can be rewritten as

$$\frac{\partial E}{\partial x}=\mu\frac{\partial^2 u}{\partial y^2}-\rho\frac{\partial u}{\partial t} \tag{11}$$

In unsteady flow, it should be noticed that all the three terms in above equation are functions of time. Since the streamwise velocity in pressure driven flow is produced by the gradient of the total mechanical energy from the Axiom (1), Eq.(11) shows the relation of velocity dependence on temporal and spatial spaces. It can be found that the streamwise velocity comes from the contributions of two terms, one is from the gradient of the viscous shear stress and another is from the time derivative of the velocity. This is an implicit relation for the streamwise velocity.

At the inflection point A, $u''=0$ (i.e. $\frac{\partial^2 u}{\partial y^2}$=0) at a moment, the Eq.(11) becomes,

$$\frac{\partial E}{\partial x} = -\rho \frac{\partial u}{\partial t} \tag{12}$$

It is seen from Eq.(12) that the velocity at the inflection point is simply produced by the time derivative of the velocity. In a period of unsteady flow, the velocity variation may be divided into the following three stages of $\frac{\partial u}{\partial t} > 0$, $\frac{\partial u}{\partial t} = 0$, and $\frac{\partial u}{\partial t} < 0$. The discussion on the velocity at the inflection point will be carried out for these three stages for the velocity profile showing $u''=0$ at *A* as in Fig.5 (b).

(1) When $\frac{\partial u}{\partial t} > 0$, we obtain $\frac{\partial E}{\partial x} < 0$ from Eq.(12). According to the discussion about Axiom (1), $u_A > 0$ is obtained at the inflection point. Since the value of $\left|\frac{\partial u}{\partial t}\right|$ is very small, both the values of $\left|\frac{\partial E}{\partial x}\right|$ and $|u_A|$ are very small, and $u_A \approx 0$.

(2) When $\frac{\partial u}{\partial t} = 0$, we obtain $\frac{\partial E}{\partial x} = 0$ from Eq.(12). According to Axiom (1), $u_A = 0$ is obtained at the inflection point.

(3) When $\frac{\partial u}{\partial t} < 0$, we obtain $\frac{\partial E}{\partial x} > 0$ from Eq.(12). According to the discussion about Axiom (1), $u_A < 0$ is obtained at the inflection point. Since the value of $\left|\frac{\partial u}{\partial t}\right|$ is very small, both the values of $\left|\frac{\partial E}{\partial x}\right|$ and $|u_A|$ are very small, and $u_A \approx 0$.

Therefore, in a period of unsteady flow, if the velocity profile shows an inflection point $u''=0$ as in Fig.5 (b), the velocity at the inflection point is,

$$u_A = 0 \quad \text{or} \quad u_A \approx 0. \tag{13}$$

***Thus, when the incoming flow reaches the inflection point, the velocity suddenly changes to near zero. This means that velocity discontinuity appears at*** $y = y_A$***, which makes the Navier-Stokes equations be singular.***

It should be pointed out that the results of Eqs.(10) and (13) as well as above discussion are obtained under the condition of the system expressed by Eq.(3), as shown in Fig.3 in laboratory system. Equations (10) and (13) are not of property of Galilean invariance. If we put the system of Eq.(3) into a set of coordinates which is moving at a uniform speed of *U* along *x* direction, the velocity of fluid in the new system will be $u+U$, and the velocity at the inflection point A will be *U*. As such, in the new system, the velocity at point A is *U*, but not zero. However, the discontinuity still exists at point *A* in the new system, which is still a singularity point of Navier-Stokes equations.

The relationship between the velocity and the second derivative of the velocity can be simply analyzed by the following reasoning.

For the steady laminar flow in plane Poiseuille flow configuration as shown in Fig.3, the velocity distribution along the *y* direction has an analytical solution [37-38],

$$u = -\frac{\partial p}{\partial x}\frac{h^2}{2\mu}\left(1-\frac{y^2}{h^2}\right). \tag{14}$$

For given fluid viscosity and a position in the flow field, the streamwise velocity is proportional to the negative pressure gradient,

$$u \propto -\frac{\partial p}{\partial x}. \tag{15}$$

The pressure gradient is related to the second derivative of the velocity as follow [37-38],

$$\frac{\partial p}{\partial x} = \mu\frac{\partial^2 u}{\partial y^2} \tag{16}$$

Therefore, we obtain Eq.(17) from Eqs.(15) and (16) at a specified position in the flow field for a given fluid,

$$u \propto -\frac{\partial^2 u}{\partial y^2}. \tag{17}$$

Although Eq.(14) is not exactly established near an inflection point of velocity profile in unsteady flow, the tendency of the velocity variation with the value of $\frac{\partial^2 u}{\partial y^2}$ in Eq.(17) should be correct. Therefore, as seen in Eq.(17), ***the value of velocity at an inflection point*** ***(*** $\frac{\partial^2 u}{\partial y^2}$ ***=0) is clearly about zero.*** This analytical result is consistent with Eqs.(10) and (13). As such, the velocity discontinuity always exists at an inflection point no matter how large of the value of the incoming velocity. However, it should be emphasized that Eq.(17) is only correct for pressure drive flows, not for shear driven flows.

In Fig.5, with further flow development, the velocity profile in Fig.5(b) may evolve into the one as shown in Fig.5(c), where a section of $u''>0$ is produced on the velocity profile. This is typical for pressure driven parallel flows as observed from experiments and simulations [11-16].

Now, we go back to Eqs.(6) and (7). When the incoming flow approaches the inflection point, the value of $\partial^2 u/\partial x^2$ in Eq.(6) may not be zero, but its magnitude is very small compared to $\partial^2 u/\partial y^2$. There may be two cases of $\partial^2 u/\partial x^2>0$ and $\partial^2 u/\partial x^2<0$ at the inflection point *A*.

(1) $\partial^2 u/\partial x^2>0$ at point *A*. Looking at Fig.5(b), it is seen that $\partial^2 u/\partial y^2<0$ on the velocity profile except at the inflection point *A*. Since the magnitude of $\partial^2 u/\partial x^2$ is very small, a minute displacement upward or downward from the inflection point *A* is able to make $(\partial^2 u/\partial x^2+\partial^2 u/\partial y^2)=0$. Thus, there is always one position near point *A* which makes Eq.(7) established.

(2) $\partial^2 u/\partial x^2<0$ at point *A*. Looking at Fig.5(c), a minute displacement downward from the inflection point *A* is able to make $(\partial^2 u/\partial x^2+\partial^2 u/\partial y^2)=0$, where $\partial^2 u/\partial y^2>0$ and it can offset the value of $\partial^2 u/\partial x^2$ since the magnitude of $\partial^2 u/\partial x^2$ is very small compared to $\partial^2 u/\partial y^2$.

Therefore, even if the value of $\partial^2 u/\partial x^2$ is not zero at the inflection point *A*, Eq. (7) is still correctly established in the neighborhood of the point *A*. There is always one point (near point *A*) on the velocity profile where $(\partial^2 u/\partial x^2+\partial^2 u/\partial y^2)=0$ is true after the velocity inflection point appears. Nevertheless, it is required for the case of $\partial^2 u/\partial x^2<0$ at point *A* that a section of $\partial^2 u/\partial y^2>0$ on the velocity profile is formed as shown in Fig.5(c). As such,

Eqs.(8) and (11) are established under these conditions. Finally, it is concluded that the neglected term $\partial^2 u/\partial x^2$ from Eq.(6) to Eq.(7) does not affect the results of Eqs.(10) and (13).

Therefore, we obtain the Theorem 2.1 according to the derivation from Eq.(3) to Eq.(17):

**Theorem 2.1. The inflection point on the velocity profile is a singularity point of the Navier-Stokes equations, where velocity discontinuity occurs.**

In the case of an inflection point appearing at the wall, as shown in Fig.4(b), the velocity at the inflection point always equals to the wall velocity, $u=0$. Thus, the problem of discontinuity of velocity does not exist in this case.

In the case of an inflection point appearing not at the wall, but $u''>0$ near the wall, as shown in Fig.4(c), this case only occurs in flows with adverse pressure gradient, for example, in divergent channel flows [37-38]. This case of non-parallel flows is not discussed in this study.

Thus far, the singularity of Navier-Stokes equations is theoretically discovered through the discussions on Eq.(3) to Eq.(17), and discontinuity of the streamwise velocity must occur at an inflection point of velocity profile or its neighborhood.

## 3 Existence and Smoothness of the Solution of the Navier-Stokes Equations

Following the discovery of the velocity discontinuity in previous section, we give some discussions on the existence and smoothness of the solution of the Navier–Stokes equations proposed in [35]. For the flow between two parallel plane plates studied here, for the given initial smooth velocity distribution of laminar flow in three-dimensional spaces (Fig.3), with the flow development, the interaction of the base flow with disturbances may lead to the velocity profile distorted at a sufficient high Reynolds number. The second derivative of the velocity with the direction normal to wall is subjected to change with time and spaces. If its magnitude is reduced to zero somewhere, the appearance of velocity inflection is not avoidable. This has been observed in experiments and simulations [11, 15, 17, 18].

Nishioka et al did detailed experiment for the plane Poiseuille flow (flow between two parallel plane plates) [11]. At the Reynolds number about 7200, a smooth laminar velocity distribution evolves into a velocity profile with inflection point in downstream under the disturbance interaction. Luo et al simulated the plane Poiseuille flow with DNS method [15]. The initial flow is a smooth parabolic laminar velocity with linear disturbance waves. They found that the velocity profile is gradually modified with time, and inflectional velocity distribution is formed during the transition of laminar flow to turbulence. In the experiments and simulations of pipe Poiseuille flows, it is also observed that a smooth laminar velocity profile is able to be developed as inflectional velocity profile before the turbulent transition occurs [13,14,16]. In summary, during the transition of laminar flow to turbulence in both plane and pipe Poiseuille flows, the velocity profile must undergo the stage of evolution of the inflectional velocity distribution. This is the sole approach for turbulent transition in both plane and pipe Poiseuille flows.

At a relative low Reynolds number, the base flow may not be affected by the disturbance and finally the laminar flow is kept [25, 27]. In this case, the disturbance is damped and inflection point is not produced finally, and the smooth laminar velocity profile is obtained. For plane Poiseuille flow, the minimum critical Reynolds number for turbulent transition is about 1000 ( $\mathrm{Re}=\frac{u_0 h}{\nu}$ , $u_0$ is the centerline velocity of the channel and $h$ is the half width of the channel between the plates in Fig.3), as summarized in [25, 27]. At a relative high Reynolds number larger than 1000, the base flow is affected by the disturbance and finally inflection point may be generated on the velocity profile depending on the Reynolds number and the disturbance. In this case, the disturbance is amplified and the base flow profile is modified gradually [15]. If both the Reynolds number and the disturbance are sufficient large, finally turbulent transition may take place. When an inflection point is produced on the velocity profile, velocity discontinuity occurs according to Eq.(8) to Eq.(17). Since the velocity is discontinuous at inflection point, the first-order and the second-order derivatives of the velocity do not exist. ***This leads to the velocity not differentiable around this position.*** As such, smooth and physically reasonable solution of Navier-Stokes equations does not exist in the flow field.

Therefore, we obtain the Theorem 3.1:

**Theorem 3.1. There exist no smooth and physically reasonable solutions of Navier-Stokes equations at high Reynolds number (beyond laminar flow).**

## 4 Discussions

### 4.1 Comparison of Theory with Experiments and Simulations on Transition

The proposed theory can be compared with experiments and simulations of laminar-turbulent transition. For pressure driven laminar flows of incompressible fluid in parallel flow configurations, we have obtained in the previous section that the inflection point on the velocity profile is a singular point of the Navier-Stokes equations. No matter how large of the magnitude of the incoming velocity, the velocity must change to zero theoretically at the inflection point in the laboratory system. However, owing to the resistance produced by the fluid viscosity, the streamwise velocity is actually not zero there, but a deep valley (or negative spike) is produced (Fig.6a). Simultaneously, a positive spike of velocity normal to the wall is produced due to the mass continuity. Since this discontinuity happens in parallel flow and the discontinuity width is theoretically zero, its occurrence is abrupt. This means that a nonlinear instability with catastrophe occurs. At the discontinuity, pulse of pressure may be generated at the inflection point owing to the conservation of the total mechanical energy (Fig.6b). On the other hand, the value of the velocity after this discontinuity may be indefinite owing to the behavior of singularity. This may be the reason why the behavior of turbulence is generally indefinite for a given laminar base flow and boundary conditions.

According to the principle of singularity of Navier Stokes equations, the mechanism of turbulent transition may be explained as follow. When the incoming laminar flow reaches a position where an inflection point exists on the velocity profile, discontinuity of velocity is generated and the streamwise velocity shows sharp negative spike. This discontinuity leads to generation of velocity normal to the wall due to the mass continuity. Simultaneously, pressure pulse is produced owing to the energy conservation. This pressure pulse may be the main origin of pressure fluctuation in wall region. At a distance normal to the wall where the inflection point locates, high-shear layer is formed due to large velocity gradient. The high-shear layer interacts with the spike leading to velocity and pressure fluctuations. On the other hand, the velocity inflection results in the formation of spanwise vortex, which rolls up downstream and then combines the streamwise vortex pair to develop as hairpin vortex. Then, the hairpin vortex and its induced younger hairpin vortices evolve into vortex packet [21, 41, 42], and finally, the vortex packet breaks down into a turbulent spot. Further, the turbulent spot increases in size downstream and the flow may spread into turbulence at a sufficient high Reynolds number.

Nishioka et al. did measurement of instantaneous streamwise velocity for plane Poiseuille flow at Reynolds number of 5000 [21]. Figure 7 shows the velocity temporal recording at a wall-normal position y/h=0.51 for the one-spike stage. The excitation of disturbance is introduced at the upstream, then it is interacted with the base flow and the velocity profile is distorted. As the increase of the disturbance amplitude, negative spike appears periodically on the velocity distribution. This phenomenon was referred as secondary instability in [20, 21]. As the disturbance amplitude increases further, the first spike evolves into two spikes, three spikes, and five spikes in each period. Finally, an irregular higher-frequency fluctuation is produced, which was called as a tertiary instability in [20, 21]. In the stage of five spikes, the maximum amplitude of the spike is located at y/h=0.38 normal to the wall. It can be found from Fig.7 that the appearance of the negative spike on the streamwise velocity agrees well qualitatively with the model proposed in Fig.6(a). These negative spikes in Fig.7 are the results of the disturbance interaction with the streamwise velocity profile as observed by the period of the spike.

Schlatter et al. did some large-eddy simulation (LES) and DNS of spatial transition for plane channel flow (plane Poiseuille flow) [24]. In the stage of nonlinear disturbance development, the velocity profile shows distinct spanwise “peak-valley splitting” of the fluctuation amplitudes. A typical velocity profile is taken at a wall-normal position $y/h$=-0.47 in the fluctuation peak plane (about the spanwise direction) with associated with the high-shear layer, as shown in Fig.8. Here, $y/h$=0 expresses the center plane between the two parallel walls and $y/h$=$\pm$1 refers to the upper and down walls respectively. It is found that a typical low-velocity “spike” exists in the velocity distribution along the streamwise direction. Then, the one spike evolves into two spikes, three spikes, four spikes, and multiple spikes downstream [24]. It is seen from Fig.8 that this one “spike” in streamwise velocity distribution is qualitatively in agreement with the valley model proposed in Fig.6(a).

Therefore, the valley model of velocity in terms of discontinuity of Navier-Stokes equations is confirmed. It is also noticed from Fig.8 that the amplitude of the spike is about 50% of the streamwise velocity. Since it is observed from the simulation result that this spike is the first one and there is no hairpin vortex in the flow, this large amplitude of negative spike is not induced by hairpin vortex.

Han et al did an experimental investigation of a late stage of transition in a circular pipe flow (pipe Poiseuille flow) subjected to periodic perturbations emanating from the wall [43]. The experimental results showed that the breakdown to turbulence starts with the appearance of spikes (velocity valley) in the temporal traces of the streamwise velocity. The transition is featured by spikes, high-shear layer, and gradually lifting-up hairpin vortices. In the upstream, one spike appears in a wave period, then the spike is doubled and further tripled downstream. ***In particular, the appearance of spike is shown to be related to the inflection of the instantaneous velocity profile along the high-shear layer.*** It was concluded that there is a similarity between the late stages of the transition in boundary layer flows and in circular pipe flows. Their results indicate that the detected spikes in the velocity recording with time are similar to those in boundary layer flow [9] and plane Poiseuille flow [21], and therefore, are in agreement with the proposed valley model in Fig.6(a).

Avila et al did some experimental work of turbulent transition for pipe flow (pipe Poiseuille flow) [44]. Figure 9 shows the experimental result of pressure distribution in a puff of pipe flow during the turbulent transition. It can be found that a pressure pulse is produced at the location of the puff. For this type of pressure distribution, there has not been reasonable explanation in the literature. It can be observed that this distribution of pressure is in agreement qualitatively with the theoretical result in Fig.6(b), in which a discontinuity of streamwise velocity leads to the pulse of pressure. It is believed that the strong fluctuation of pressure in wall bounded turbulence is related to the pressure pulse generated by discontinuity of streamwise velocity.

The phenomenon with high pressure pulse is also found in the transitional experiment of high-speed boundary layer flow [45]. A region of high pressure is detected beneath the instability wave packets which evolve into turbulent spots. At the position where an arrowhead shape of the wave packet is formed, the value of a peak pressure reaches about 40% above the freestream pressure. Thus, we have reason to believe that the "bursting events" in turbulence is associated with the high pressure pulse (zone) resulted from the discontinuity of streamwise velocity.

Nepomuceno and Lueptow made pressure and shear stress measurements at the wall in a turbulent boundary layer of axial flow on a cylinder [46], and found a positive peak in the wall pressure near the wall. Then, they suggested that a positive wall pressure peak event is the key wall pressure signature associated with the burst cycle.

In the experiments and simulations, with the flow downstream or with the increasing Re number, or with the increasing disturbance amplitude, it is found that the one spike stage formed at the beginning evolves into two spikes, triple spikes, four spikes, or multiple spikes [21-24]. The mechanism for these processes can be explained by the fluctuation/oscillatory variation of the velocity profile with the flow forward. Owing to the interaction of the velocity discontinuity and the high-shear layer, the flow is unstable locally. When the inflectional velocity profile is produced, negative velocity spike appears. Since there is no adverse pressure gradient in the boundary layer, the mean flow is not able to flow reversely. Then, the inflection point disappears and the instantaneous velocity profile becomes normal. Further, due to the action of the disturbance and the high-shear layer, the velocity inflection appears again. In such away, multiple spikes are generated, as shown in Fig.10. .

Dou and co-authors have reached the conclusion before by employing the energy gradient theory that the inflection point on the velocity profile is a singularity of the Navier-Stokes equations in pressure driven flows [25, 27, 29]. At the inflection point, the energy gradient function $K$, which can be considered as a ***local Reynolds number***, is infinite. As such, even if with a very small amplitude of disturbance, instability can be stimulated.

As early as in 1880, Rayleigh obtained by linear stability analysis that the necessary condition for instability of inviscid flow is the existence of an inflection point on the velocity profile [47]. Later, Tollmien proved that the inflection point on the velocity profile is also a sufficient condition for instability of inviscid parallel flows with a symmetric profile [48].

It needs to be pointed out that the physical mechanism of the inflection point in inviscid flow is completely different from that of the inflection point in viscous flow. In inviscid flow, there is no discontinuity of streamwise velocity at the inflection point. In viscous flow, there is discontinuity of streamwise velocity at the inflection point. In inviscid flow, there is no loss of the total mechanical energy along the streamline in the whole flow field. In viscous flow, there is a drop of the total mechanical energy along the streamline, except at the inflection point where the effect of viscosity does not appear due to $u''=0$.

For laminar flow of plane Couette flow configuration, $u''=0$ everywhere in the whole flow field. This does not mean that viscosity does not play a role, because input of external work counteracts exactly the energy loss due to viscosity along a streamline. Thus, there is no discontinuity at a position with $u''=0$ [28].

## 4.2 Burst Event in Laminar-Turbulent Flow Transition

In wall bounded turbulence, experiments and simulations have shown that the turbulent motion near the wall is featured by a burst event [10, 23, 41]. It was suggested that the bursting process plays a crucial role in turbulence production and in the interaction between the inner and outer layers [10, 41]. However, in the past 50 years or so, the physical mechanism of the burst event ("eruption" or "breakup") found in experiments is still not fully understood.

The usage of the word "burst" (or "bursting") in the public literature is confusing. The original statement in Kline et al. [10] means that "burst" is an eruption of the lifting of the low-speed fluid near the wall (the streamwise vorticity "lift the streaks gradually away from the wall until a point is reached where some sort of sudden instability appears to occur"). In some late studies, the word "burst" is explained as the cycle of the ejection-sweep [41, 42]. Generally, the "ejection" and "sweep" express the motions of fluctuation in second and fourth quadrants, respectively. In near wall turbulence, the cycle of ejection-sweep always exists due to the role of streamwise vortex pair, but differing in the magnitude everywhere. The interpretation of cycle of ejection and sweep is obviously different from the meaning of "burst" in Kline et al [10]. As summarized in Robinson [41], there are two general concepts about "burst" mentioned. In present study, we prefer to employ the first definition of burst: "a violent, temporally intermittent eruption of fluid away from the wall, in which a form of local instability is often implied."

The discontinuity of streamwise velocity predicted by the present theory (the valley model in Fig.6(a)) is consistent with the "burst" phenomenon found from experiment. They share the following features:

(1) Local nonlinear instability occurs;

(2) Negative spike of streamwise velocity;

(3) Low-speed fluid temporally lifts up, meaning positive spike of velocity normal to wall;

(4) It occurs suddenly and explosively.

Thus, it can be judged that the sharp negative spike induced by the velocity discontinuity in Fig.6(a) is the "burst" phenomenon found in experiment of turbulent transition, which occurs under the condition of inflectional velocity profile [10,23,41]. Actually, in some researches, the negative velocity spike is referred to as burst, for example, in [23]. Within one burst, there may be one or more spike/ejection structures [23].

At a sufficient high Reynolds number laminar flow, the incoming velocity profile interacts with three-dimensional disturbance leads to formation of streamwise vortex pairs near the wall. The streamwise vortex pair near the wall feeds the fluid up and down and forms low-speed streaks and shear layer periodically along the spanwise direction. In the downstream, the shear layer gradually lifts up along streamwise direction. At a sufficient distance downstream, the streamwise velocity profile along the shear layer becomes inflectional. At this position, nonlinear instability first occurs due to the most inflectional velocity profile, and streamwise velocity discontinuity appears. Further, the first negative spike (or burst) of velocity is generated which leads to local high pressure. In such away, low-speed fluid near the shear layer erupts to lift-up (leads to high shear stress). This description is consistent with the burst phenomenon observed in the channel and boundary layer experiments [10, 23]. Simultaneously, spanwise vortex will be formed at the inflection point position. Then, the spanwise vortex will combine the streamwise vortex pair to form the hairpin vortex downstream.

In the transition of a laminar flow to turbulence, ***the "burst" event is a critical phenomenon of turbulent transition***, which symbolizes the beginning of nonlinear instability. The energy of "eruption" of the bursting phenomenon is coming from the local high pressure caused by the velocity discontinuity. This phenomenon can only be produced by three-dimensional disturbance, and the location of the first spike is usually occurring at the position under the tip of the $\Lambda$ wave (if it appears) where the velocity profile is the most inflectional and is the most dangerous to loss the stability. The first negative spike results in the first hairpin vortex. Therefore, the appearance of negative spike (burst) is a nonlinear instability rather than secondary instability. It is understood from present study that turbulent transition is completely determined by the singularity of Navier-Stokes equations due to velocity discontinuity, and the appearance of Tollmien-Schlichting wave (T-S wave) is not necessary. If a T-S wave is introduced to a laminar flow, and if the development of the T-S wave could not finally results in velocity discontinuity, turbulent transition could not occur and eventually the laminar flow would be kept (for example, flow at low Reynolds number). Oppositely, if any kind of disturbance could cause occurrence of velocity discontinuity, turbulent transition would be excited. This finding may provide with new approaches for control of turbulence.

Since the critical condition of turbulent transition is featured by the generation of the bursting phenomenon and it is completely a nonlinear instability caused by three-dimensional finite disturbance, linear stability theory is not able to predict it. The energy gradient theory describes the relative significance of the disturbance amplification and the viscous damping in dominating the flow stability [25-34]. A scaling $\mathrm{Re}^{-1}$ of the disturbance amplitude with the Reynolds number at transition was obtained theoretically for the normal disturbance to the flow [27], which obtains agreement with experiments for plane Poiseuille flow [7] and pipe Poiseuille flow [49].

### 4.3 Sustenance of Fully Developed Turbulence by Singularities

For fully developed turbulent flows, the governing equations are still the unsteady Navier-Stokes equations Eq.(3). The difference is that the incoming boundary condition is changed as full developed turbulence. The discussions for Eq.(3) to Eq.(13) are still correct for fully developed turbulent flows.

In fully developed turbulent flow, there are numerous vortices distributed in the flow field, large scales and small scales. These vortices move downstream with spatial and temporal variations. When a vortex exists in the flow, the shape of the velocity profile can be modified by the vortex. The streamwise velocity induced by a spanwise vortex is depicted as in Fig. 11. When this vortex is overlapping with the incoming velocity profile, an inflectional velocity profile is generated, as shown in Fig.12, where a point of $\partial^2 u / \partial y^2 = 0$ is shown on the velocity profile. Thus, the center of each spanwise vortex contains one center of inflectional velocity profiles, and a singularity point exists nearby the vortex center where the streamwise velocity is theoretically zero for the system of Eq.(3).

According to analysis in previous sections, the flow is always unstable at an inflection point of velocity profile due to singularity of the governing equations. When fluid particle passes this point (singularity), discontinuity of velocity is generated, and the value of velocity becomes indefinite. Meanwhile, the velocity becomes to be of oscillation/fluctuation, and pressure fluctuation is produced due to the role of interaction of high-shear layer with the velocity discontinuity. The velocity discontinuity and the resulted pressure pulse form the power to drive the fluctuation/oscillation of the flow. This is why the most instability place in the wake of a cylinder always occurs at the center of vortices. Dou and Ben did the simulation of pressure driven flow past a cylinder between two parallel walls [32]. It is shown that the centre of a vortex is the most instability location, which is in agreement with the experiments in literature. Therefore, it may be said that turbulence is composed of a large quantity of singularities and turbulence is sustained by singularities in flow field.

A large quantity of vortices in turbulence results in a flow field with fluctuations of velocity and pressure. At these locations, the incoming flow is trapped into singularities. When the fluid flow passes these singularities, the velocity becomes ***indefinite*** and the flow behavior is ***not predictable***. The velocity is not differentiable at the vortex centre. Thus, ***there exist no smooth and physically reasonable solutions of Navier-Stokes equations for fully developed turbulence***.

Now, it is well known that there are coherent structures in full developed turbulence which feature the large-scale organized structure of vortex motions composed of hairpin vortices and other types of vortices [41, 42]. These large-scale vortices are generated from the formation of the singularity of the mainstream velocity by the interaction between the mainstream velocity and disturbance as discussed previously.

The roles of these large-scale vortices include inducing smaller-scale vortices, transferring energy from outer flow to near wall flow, and sustaining turbulence by providing with the flow structures to produce "burst event," etc.:

(1)These large-scale vortices induce smaller-scale vortices at various spatial directions.

(2)These large-scale vortices transfer energy from the mainstream flow to the small-scale vortices grade by grade, and finally the energy is dissipated by small vortices. For example, the streamwise vortices bring the high-speed fluid to the near wall zone by the sweep process and lift the low-speed fluid to the outer layer by the ejection process, and the spanwise vortices transfer energy from the mainstream streamwise flow to the near wall flow by the rolling-up process.

(3)The legs of hairpin vortices (streamwise vortices) promote the formation of velocity inflection of the streamwise velocity which results in low-speed streaks and high-shear layer near the wall. The latter leads to burst occurrence (single or group of negative spikes). The vortex head (spanwise vortex) of hairpin vortices also causes discontinuity (results in negative spike) of the streamwise velocity profile by the velocity inflection, as shown in Figs.11-12.

Thus, the unstable high-shear layer and the vortex heads of the hairpin vortices play significant roles in sustaining turbulence since they provide with large scale of flow discontinuities. As such, the full developed turbulence is maintained by the large-scale vortices resulted from streamwise velocity. If the large-scale vortices are destroyed, the small-scale vortices cannot continue to obtain energy from the large-scale vortices, and the turbulence will not be maintained. Therefore, turbulence is maintained by large scale vortices formed by the mainstream flow. If these large scale vortices are destroyed, turbulence would be killed.

In the experiment on a turbulent boundary layer [50], the velocity discontinuity is clearly seen in the outer part of the boundary layer. The velocity discontinuity occurring in boundary layer flow [39-40] is similar to that occurring in the plane Poiseuille flow (or channel flow). When the fluid particle in the boundary layer reaches a point where the velocity profile is sufficiently distorted, the velocity suddenly changes to zero theoretically determined by the Navier-Stokes equations in a system similar to Fig.3. As such, streamwise velocity discontinuity takes place and a deep velocity valley (negative spike) is produced [9, 10, 19, 20]. This discontinuity results in the flow fluctuation/oscillation and the velocity to be indefinite. Since the boundary layer flow is not belong to pressure driven flow, the Axiom (1) discussed in previous section is not suitable for boundary layer flow. The detailed process of singularity generation in boundary layer flow will be given in separate publication, according to different energy transfer process following the energy gradient theory [25, 27, 29, 31].

Since similar spikes in the streamwise velocity appear in the late stages of turbulent transition for plane Poiseuille flow, pipe Poiseuille flow, and boundary layer flow as found in experiments [9, 19, 21, 43], there must be similarity for their turbulent transitional behavior and mechanism. Monty et al. compared turbulent flow in pipe, channel and boundary layer flows, and obtained that there is some similarity about the turbulent structure for these three types of flows although it has been observed large-scale differences between pipes/channels and boundary layers [51]. We believe that the singularity of Navier-Stokes equations induced by the streamwise velocity discontinuity is the sole mechanism of transition of laminar flow to turbulence, including plane Poiseuille flow, pipe Poiseuille flow, plane Couette flow, and boundary layer flow, etc..

Thus, since turbulence is dominated by the field of the streamwise velocity and it is driven by the discontinuity of the streamwise velocity, the structure of large scale turbulence is not random and is not homogeneous isotropic. In fully developed turbulence, the roles of fluctuations of velocity components $u$ and $v$ are not equal. The former is active and the latter is passive.

In previous studies [10, 20, 21, 41], it was emphasized that the burst event (negative spikes) can play important role in turbulent transition and fully developed turbulence. However, why the spike is negative, why the spike amplitude is so large, and why the occurrence of burst is suddenly, are not explained reasonably before. In present study, all of these questions have been clarified with the proposed theory.

In summary, transition of a laminar flow to turbulence is implemented via the singularity of Navier-Stokes equations, which is the sole approach for the transition. ***This mechanism is universal for turbulent transition in parallel shear flows.*** The approaches of occurrence of singularity are different in pressure driven flow and shear driven flow due to different energy transfer process. However, in both types of flows, it is the singularity of Navier-Stokes equations to lead to turbulent transition.

Now, the following questions can be answered from this study:

(1) Why can a smooth laminar flow be transited to turbulence? This is due to singularity of Navier-Stokes equations at an inflection point. The value of the velocity of fluid particle passing this position will be discontinuous and indefinite.

(2) Why is there strong oscillation/fluctuation in turbulence? This is caused by the discontinuity of flow parameters at inflection point formed by disturbances and vortices. Any magnitude of incoming velocity must drop to zero theoretically at this location in a system expressed by Eq.(3), which results in pressure pulse and flow oscillation/fluctuation.

(3) Why does large scale coherent structure coexists with small scale random motion in turbulence? The coherent structure is dominated by the singular points of the Navier-Stokes equations, while the random motion is the result of the fluctuation/oscillation of fluid around the singular points.

(4) Why is there intermittency in turbulence? Intermittency is caused by the burst events and the vortices containing singularity which pass the observation location. The locations of velocity discontinuity (inflection point) are the high fluctuation/oscillation area and the zones between these locations are the intermittent area.

(5) Why can a turbulent flow not be duplicated? This is caused by the indefinite behavior of flow due to singularity of Navier-Stokes equations at discontinuity.

(6) Is there smooth and physically reasonable solution of Navier-Stokes equations? Negative answer is given, and this is because the velocity is not differentiable at the velocity discontinuity location. Therefore, there exist no smooth and physically reasonable solutions of Navier-Stokes equations at high Reynolds number (.i.e, beyond laminar flow).

## 5 Conclusions

In our previous works [25-34], energy gradient theory is developed for flow stability and turbulent transition. In this study, the singularity of Navier-Stokes equations is discovered and the mechanisms of turbulent transition and sustenance of fully developed turbulence are explained. The following conclusions can be drawn from this study.

(1) For the pressure driven incompressible flow, the singularity of Navier-Stokes equations is exactly uncovered through the derivation of the Navier-Stokes equations and the analysis of the velocity profile. It is found that there is discontinuity of streamwise velocity at the inflection point of velocity profile in viscous parallel flows, which is singular in the flow field.

(2) The valley model of velocity (proposed in this study) due to velocity discontinuity at the inflection point of the velocity profile agrees qualitatively with the experimental and simulation results of plane Poiseuille flow at low Reynolds number (transitional stage). In addition, the local pressure pulse obtained from the theory is in agreement qualitatively with the experimental observation of pipe Poiseuille flow in literature.

(3) The singularity of the Navier-Stokes equations is the cause of turbulent transition and the inherent mechanism of sustenance of fully developed turbulence.

(4) In laminar-turbulent flow transition, the "burst" event is the critical phenomenon, which is a nonlinear instability caused by three-dimensional finite disturbance. In full developed turbulence, the "burst" event is the power to generate the fluctuations/oscillations.

(5) A laminar flow, no matter how it develops under disturbance, if it cannot make the Navier-Stokes equations singular (causing velocity discontinuity), turbulent transition cannot occur. This is also the reason why the linear stability theory failed to predict the transition of turbulence.

(6) Turbulence is mainly maintained by the largest scale vortices (coherent structure) in the flow field by forming the flow discontinuities (negative spikes), which are also responsible to transfer energy from the mainstream flow to smaller scale vortices.

(7) The finding of singularity supports the criterion of turbulent transition for parallel pressure driven flows proposed based on the energy gradient theory that the necessary and sufficient condition for turbulent transition is the existence of an inflection point on the velocity profile.

(8) Since the velocity is not differentiable at the velocity discontinuity location, there exist no smooth and physically reasonable solutions of Navier-Stokes equations at high Reynolds number (beyond laminar flow).

## Acknowledgements

The author thanks Prof. B. C. Khoo at National University of Singapore and Prof. K. Xu at Hong Kong University of Science and Technology for helpful discussions. The author also thanks the anonymous reviewers for helpful comments. This work is supported by National Natural Science Foundation of China (51579224).

## References

[1] Reynolds, O., An experimental investigation of the circumstances which determine whether the motion of water shall be direct or sinuous, and of the law of resistance in parallel channels, Phil. Trans. Roy. Soc. London A, 174, 1883, 935-982.

[2] Heisenberg, W., Uber stabilitat und turbulenz von flussigkeitsstromen, Ann Phys., Lpz. (4) 74, 1924, 577-627, On stability and turbulence of fluid flows, NACA TM-1291, 1951.

[3] Lin, C.-C., On the stability of two-dimensional parallel flows, Proceedings of National Academy of Science, 30, 1944, 316-324.

[4] Orszag, S. A., Accurate solution of the Orr-Sommerfeld stability equation, J. Fluid Mech., 50, 1971, 689-703.

[5] Orszag, S. A., Patera, A. T., Subcritical transition to turbulence in plane channel flows, Physical Review Letters, 45, 1980, 989-993.

[6] Trefethen, L. N., Trefethen, A. E., Reddy, S. C., Driscoll, T. A., Hydrodynamic stability without eigenvalues, Science, 261, 1993, 578-584.

[7] Lemoult, G., Aider, J.-L., Wesfreid, J. E., Experimental scaling law for the subcritical transition to turbulence in plane Poiseuille flow, Physical Review E, 85(2), 2012, No. 025303.

[8] Theodorsen, T., Mechanism of turbulence. In Proceedings of Second Midwestern Conference on Fluid Mechanics, Ohio State University, Columbus, OH, 1952, pp. 1-19.

[9] Klebanoff, P. S., Tidstrom, K. D., Sargent, L. M., The three-dimensional nature of boundary layer instability, J. Fluid Mech., 12(1), 1962, 1-34.

[10] Kline, S.J., Reynolds, W.C., Schraub, F.A., Runstadler, P.W., The structure of turbulent boundary layers, J. Fluid Mech., 30, 1967, 741-773.

[11] Nishioka, M., Iida, S., Ichikawa, Y., An experimental investigation of the stability of plane Poiseuille flow, J. Fluid Mech., 72, 1975, 731-751.

[12] Hof, B., van Doorne, C. W. H., Westerweel, J., Nieuwstadt, F. T. M., Faisst, H., Eckhardt, B., Wedin, H., Kerswell, R. R., Waleffe, F., Experimental observation of nonlinear traveling waves in turbulent pipe flow, Science, 305(5690), 2004, 1594-1598.

[13] Hof, B., de Lozar, A., Avila, M., Tu, X., Schneider, T.M., Eliminating turbulence in spatially intermittent flows, Science, 327(5972), 2010, 1491-1494.

[14] Nishi, M., Unsal, B., Durst, F., Biswas, G., Laminar-to-turbulent transition of pipe flows through puffs and slugs, J. Fluid Mech., 614, 2008, 425-446.

[15] Luo, J., Wang, X., and Zhou, H., Inherent mechanism of breakdown in laminar-turbulent transition of plane channel flows, Science in China Ser. G Physics, Mechanics and Astronomy, 48(2), 2005, 228-236.
[16] Wedin, H., Kerswell, R. R., Exact coherent structures in pipe flow: travelling wave solutions, J. Fluid Mech., 508, 2004, 333-371.
[17] Biringen, S., Final stages of transition to turbulence in plane channel flow, J. Fluid Mech., 148, 1984, 413-442.
[18] Sandham, N. D., Kleiser, L., The late stages of transition to turbulence in channel flow, J. Fluid Mech., 245, 1992, 319-348.
[19] Borodulin, V. I., Kachanov, Y. S., Roschektayev, P., Experimental detection of deterministic turbulence, Journal of Turbulence, 12, 2011, Article N23.
[20] Kachanov, Y. S., Physical mechanisms of laminar-boundary-layer transition, Annu. Rev. Fluid Mech., 26, 1994, 411-482.
[21] Nishioka, M., Asai, M., Iida, S., Wall phenomena in the final stage of transition to turbulence, In: Transition and Turbulence, Ed: R. E. Meyer, Academic Press, 1981, New York, pp.113-126.
[22] Kao, T. W., Park, C., Experimental investigations of the stability of channel flows Part 1: Flow of a single liquid in a rectangular channel, J. Fluid Mech., 43(1), 1970, 145-164.
[23] Luchiktand, T. S., Tiederman, W. G., Timescale and structure of ejections and bursts in turbulent channel flows, J. Fluid Mech., 174, 1987, 524-552.
[24] Schlatter, P., Stolz, S., Kleiser, L., Large-eddy simulation of spatial transition in plane channel flow, Journal of Turbulence, 7(1), 2006, 1- 24.
[25] Dou, H.-S., Mechanism of flow instability and transition to turbulence, Inter. J. Non-Linear Mech., 41 (4), 2006, 512-517.
https://www.researchgate.net/publication/245215903
[26] Dou, H.-S., Khoo, B.C., Yeo, K.S., Instability of Taylor-Couette flow between concentric rotating cylinders, Inter. J. of Therm. Sci., 47, 2008, 1422-1435.
https://arxiv.org/abs/physics/0502069
[27] Dou, H.-S., Physics of flow instability and turbulent transition in shear flows, Inter. J. Phys. Sci., 6(6), 2011, 1411-1425. http://arxiv.org/abs/physics/0607004
[28] Dou, H.-S., Khoo, B.C., Investigation of turbulent transition in plane Couette flows using energy gradient method, Advances in Appl. Math. and Mech., 3(2), 2011, 165-180.
http://arxiv.org/abs/nlin.CD/0501048
[29] Dou, H.-S., Khoo, B.C., Criteria of turbulent transition in parallel flows, Modern Physics Letters B, 24(13), 2010, 1437-1440. https://arxiv.org/abs/0906.0417
[30] Dou, H.-S., Jiang, G., Numerical simulation of flow instability and heat transfer of natural convection in a differentially heated cavity, Int. J. Heat Mass Tran., 103, 2016, 370-381. https://arxiv.org/abs/1808.05628
[31] Dou, H.-S., Xu, W., Khoo, B. C., Stability of boundary layer flow based on energy gradient theory, Modern Physics Letters B, 32 (12), 2018, No. 1840003.
https://arxiv.org/abs/1806.07058
[32] Dou, H.-S., Ben, A. Q., Simulation and instability investigation of the flow around a cylinder between two parallel walls, Journal of Thermal Science, 24(2), 2015, 140-148.
https://arxiv.org/abs/1902.02460

[33] Xiao, M., Dou, H.-S., Wu, C., Zhu, Z., Zhao, X., Chen, S., Chen, H., Wei, Y., Analysis of vortex breakdown in an enclosed cylinder based on the energy gradient theory, European Journal of Mechanics / B Fluids, 71 (2018), 66-76. https://arxiv.org/abs/1808.05025

[34] Dou, H.-S., Phan-Thien, N., Viscoelastic flows around a confined cylinder: instability and velocity inflection, Chem. Eng. Sci., 62, 2007, 3909-3929. https://www.researchgate.net/publication/244116911

[35] Fefferman, C. L., Existence and smoothness of the Navier-Stokes equation, Clay Mathematics Institute, 2000, pp.1-6. http://www.claymath.org/millennium-problems/navier-stokes-equation

[36] Doering, C. R., The 3D Navier-Stokes problem, Annu. Rev. Fluid Mech., 41, 2009, 109-128.

[37] Schlichting, H., Boundary Layer Theory, 7th Edition, 1979, Springer, Berlin.

[38] White, F.M., Viscous Fluid Flow, 2nd Edition, 1991, McGraw, New York.

[39] Robinson, S. K., Kline, S. J., Spalart, P. R., A review of quasi-coherent structures in a numerically simulated turbulent boundary layer, NASA-TM-102191, 1989.

[40] Smits, A. J., Delo, C., Self-sustaining mechanisms of wall turbulence. In: Reguera D., Rub íJ.M., Bonilla L.L. (Eds) Coherent Structures in Complex Systems. Lecture Notes in Physics, Vol 567, 2001, Springer, Berlin, Heidelberg, pp. 17-38.

[41] Robinson, S.K., Coherent motion in the turbulent boundary layer, Annu. Rev. Fluid Mech., 23, 1991, 601-639.

[42] Zhou, J., Adrian, R. J., Balachandar, S., Kendall, T. M., Mechanisms for generating coherent packets of hairpin vortices in channel flow, J. Fluid Mech., 387, 1999, 353-396.

[43] Han, G., Tumin, A., Wygnanski, I., Laminar-turbulent transition in Poiseuille pipe flow subjected to periodic perturbation emanating from the wall, Part 2. Late stage of transition, J. Fluid Mech., 419, 2000, 1-27.

[44] Avila, K., Moxey, D., de Lozar, A., Avila, M., Barkley, D., Hof, B., The onset of turbulence in pipe flow, Science, 333, 2011, 192-196.

[45] Casper, K. M., Beresh, S. J., Schneider, S. P., Pressure fluctuations beneath instability wave packets and turbulent spots in a hypersonic boundary layer, J. Fluid Mech., 756, 2014, 1058-1091.

[46] Nepomuceno, H. G., Lueptow, R. M., Pressure and shear stress measurements at the wall in a turbulent boundary layer on a cylinder, Phys. Fluids, 9 (9), 1997, 2732-2739.

[47] Rayleigh, L., On the stability or instability of certain fluid motions, Proc. Lond. Maths. Soc., 11, 1880, 57-70.

[48] Tollmien, W., Ein allgemeines kriterium der instabilitat laminarer gescgwindigkeitsverteilungen, Nachr. Wiss Fachgruppe, Gottingen, Math. Phys., 1, 1935, 79-114. Translated as, General instability criterion of laminar velocity disturbances, NACA TM-792, 1936.

[49] Hof, B., Juel, A., Mullin, T., Scaling of the turbulence transition threshold in a pipe, Phy. Rev. Lett., 91, 2003, 244502.

[50] Falco, R. E., Coherent motions in the outer regions of turbulent boundary layers, Phys. Fluids, 20 (10, II), 1977, 5124-5132.

[51] Monty, J. P., Hutchins, N., Ng, H. C. H., Marusic, I., Chong, M. S., A comparison of turbulent pipe, channel and boundary layer flows, J. Fluid Mech., 632, 2009, 431 442.

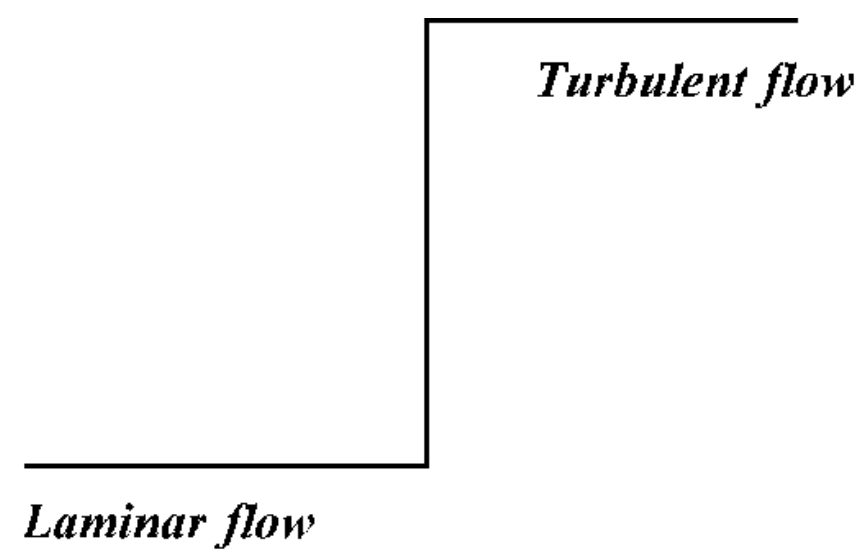

Figure 1. Discontinuity existing during transition from laminar flow to turbulent flow. There is a jump for all the flow parameters such as velocity, pressure, shear stress as well as drag force, etc.

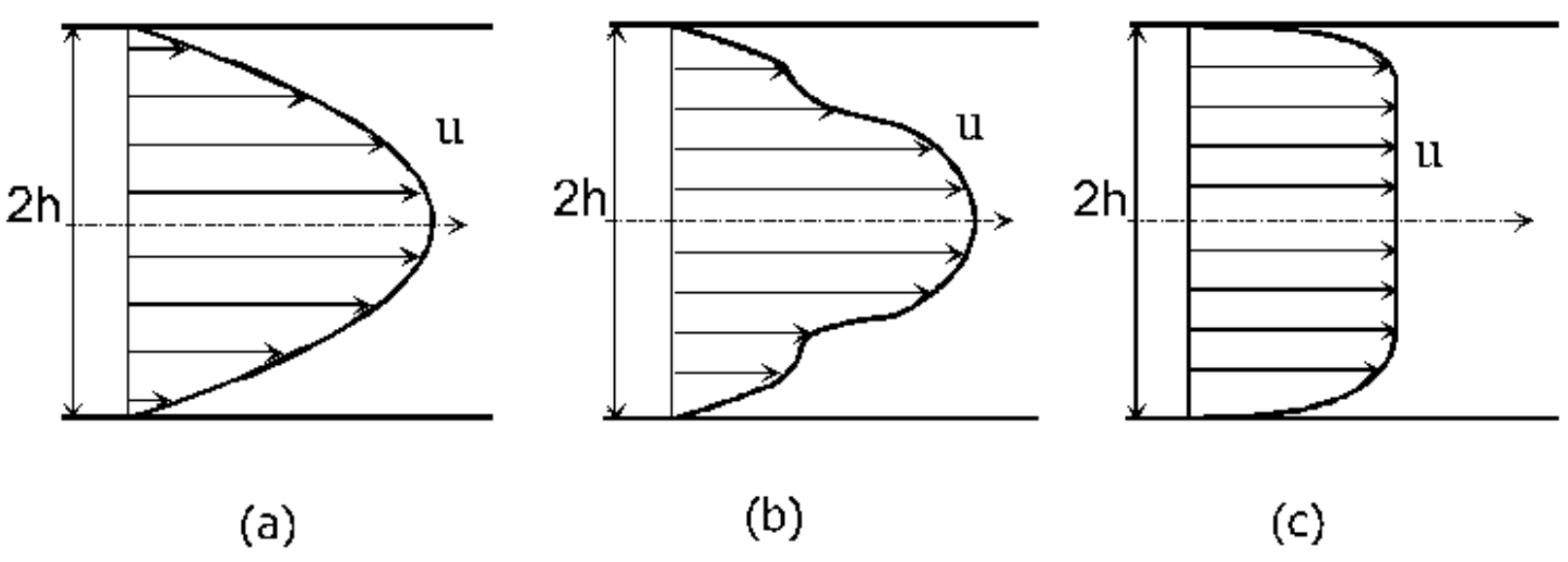

Figure 2. Velocity profile during transition from laminar flow to turbulence for plane Poiseuille flow as observed from experiments and simulations. (a) Laminar flow; (b) Transitional flow; (c) Turbulence.

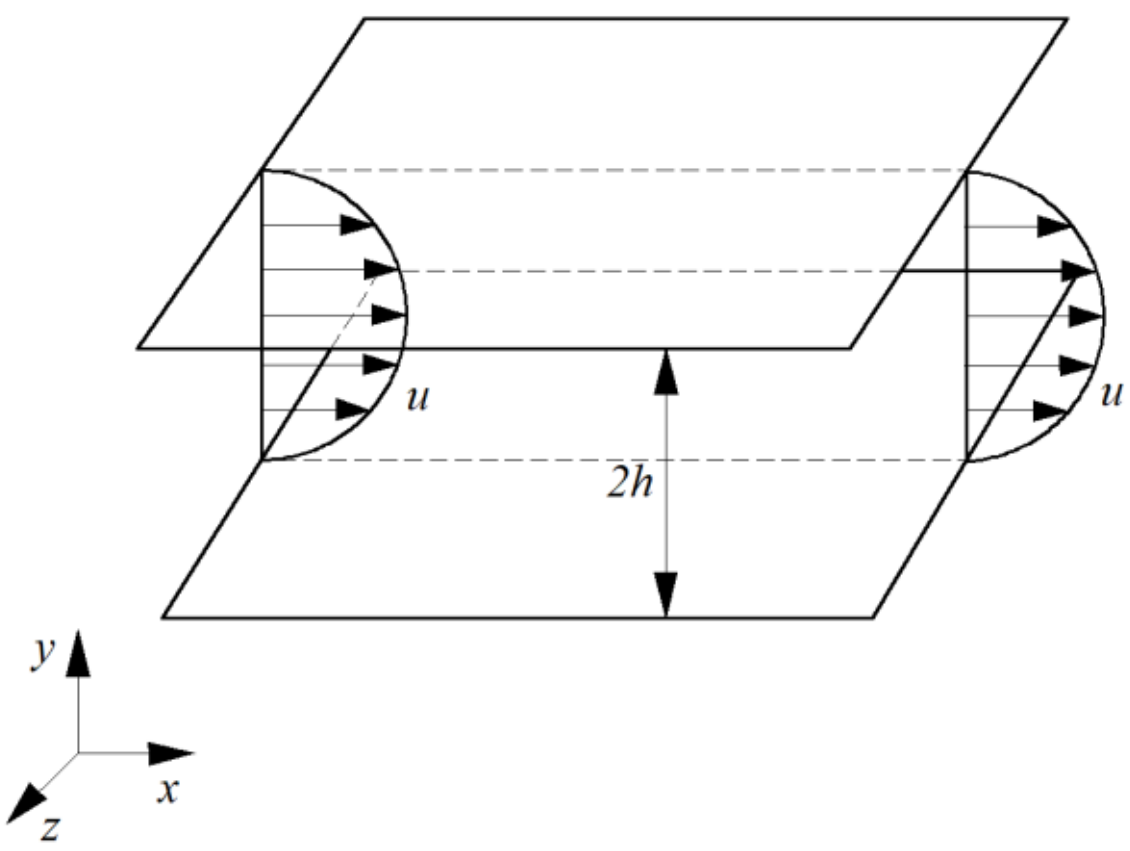

Figure 3. Plane Poiseuille flow between two parallel plates with boundary conditions and initial conditions.

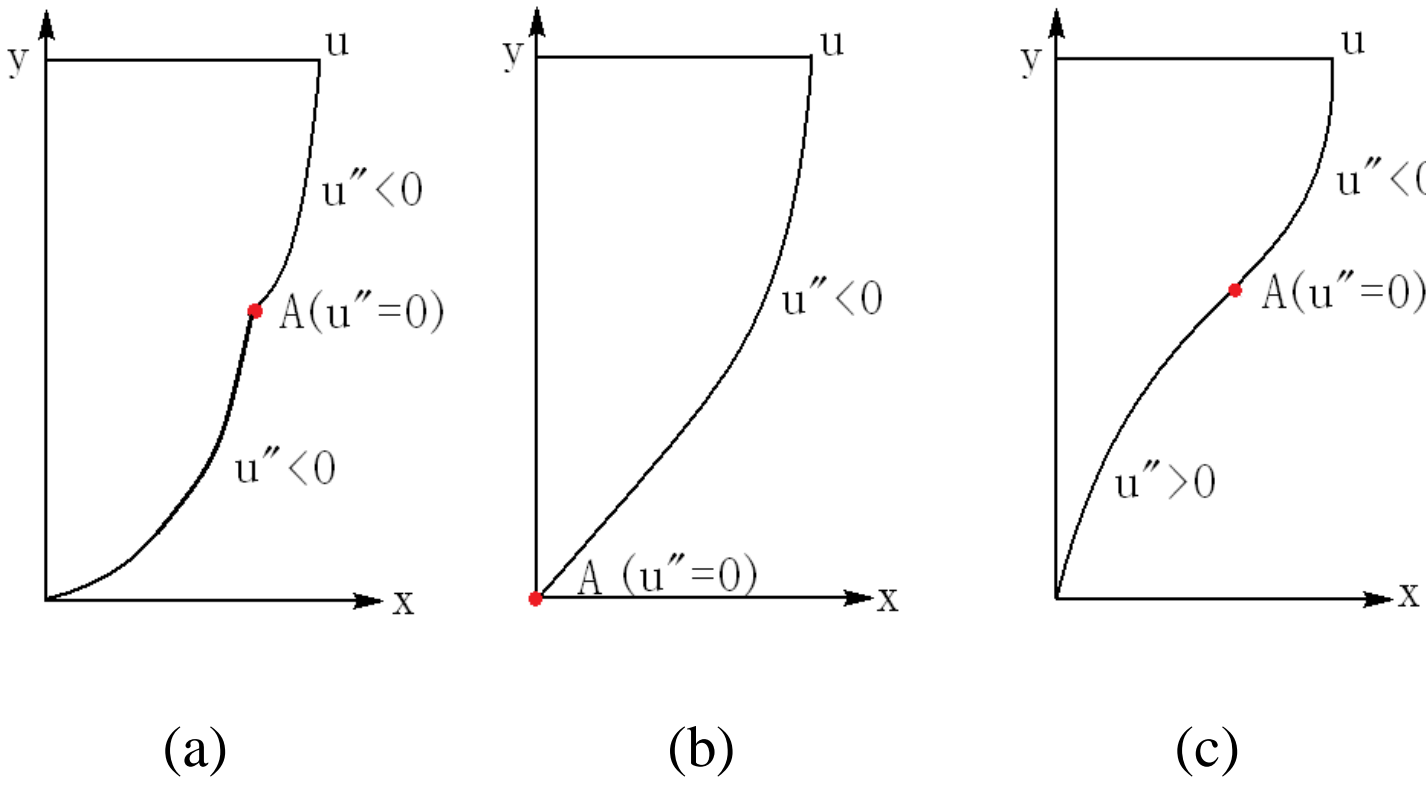

Figure 4. (a) Inflection point is not located at the wall, $u''<0$ at the wall; (b) Inflection point is located at the wall, $u''=0$ at the wall; (c) Inflection point is not located at the wall, but $u''>0$ at the wall.

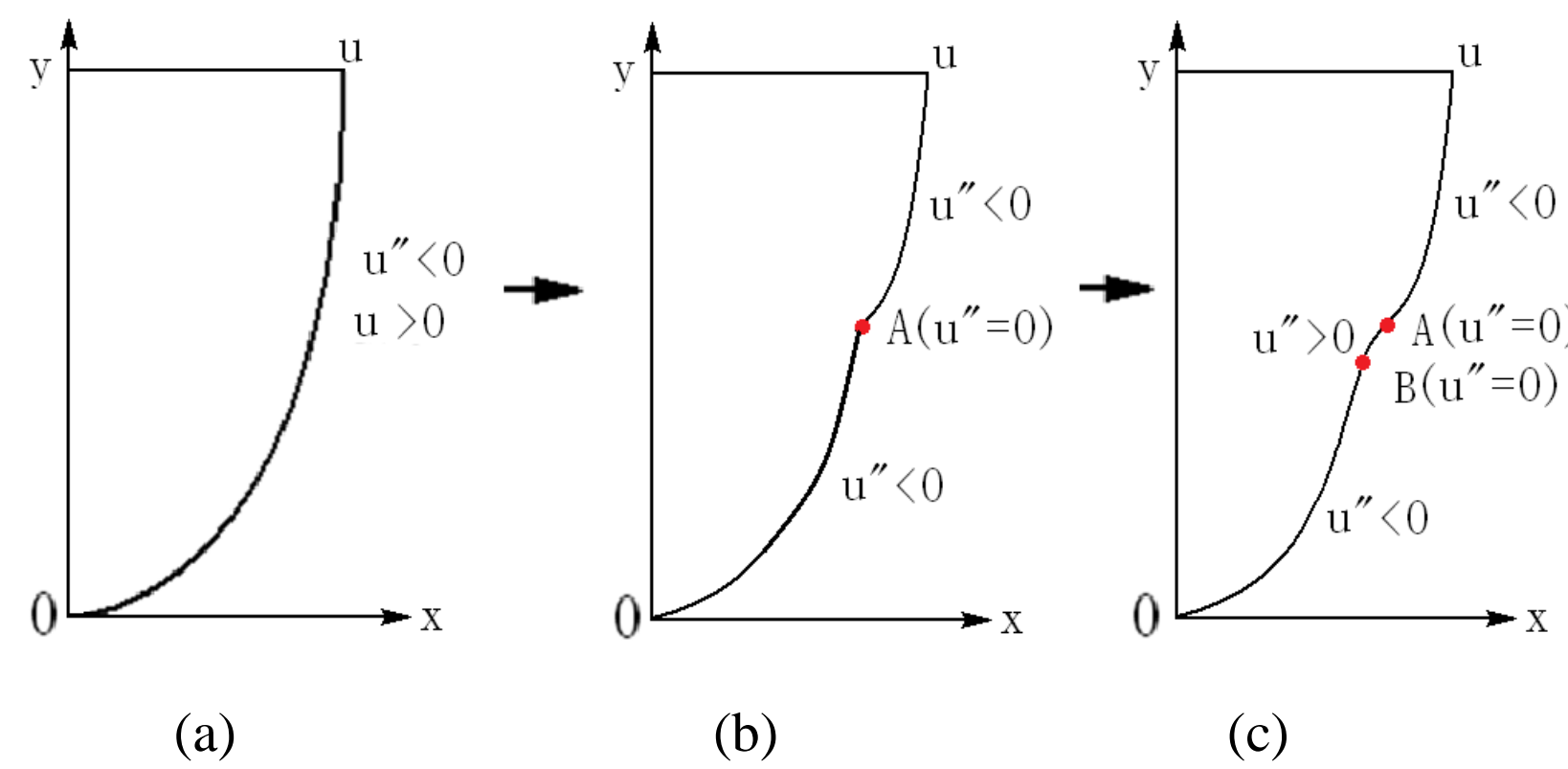

Figure 5. (a) Velocity profile of laminar flow; (b) Inflection point appears on the velocity profile; (c) A section of $u''>0$ appears on the velocity profile, and the second inflection point *B* is produced. This evolution is typical for pressure driven parallel flows [11-16].

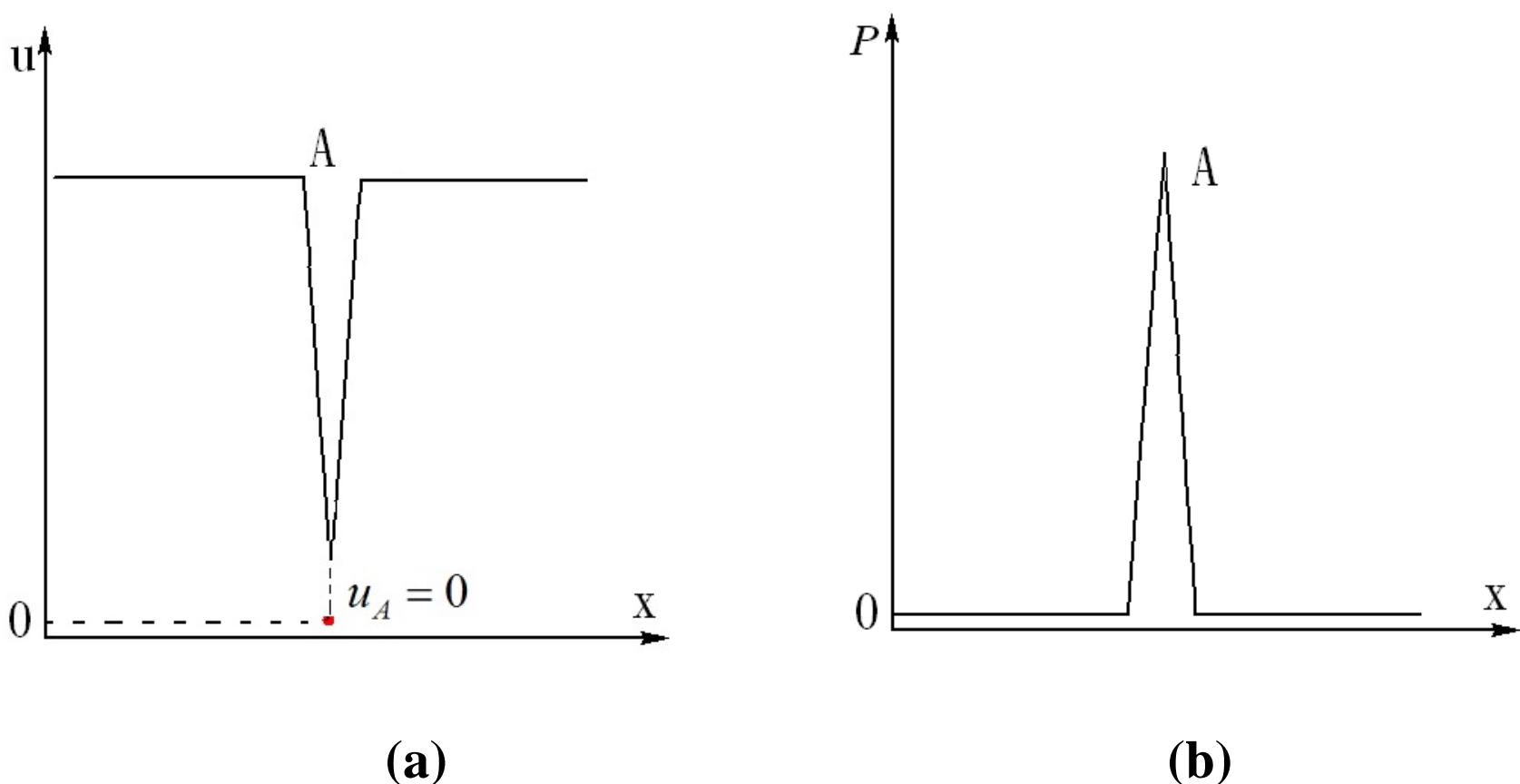

**(a)** **(b)**

Figure 6. Schematic model of streamwise velocity along the streamwise direction before and after the inflection point *A*. The incoming flow is laminar. Discontinuity occurs at the inflection point *A* with $u_A = 0$. (a) Streamwise velocity showing a velocity valley at the discontinuity. (b) Pressure distribution showing a pulse at the discontinuity owing to the conservation of total mechanical energy along a streamline. The value of velocity after the discontinuity may be indefinite, and thus the value of pressure after the discontinuity may be also indefinite.

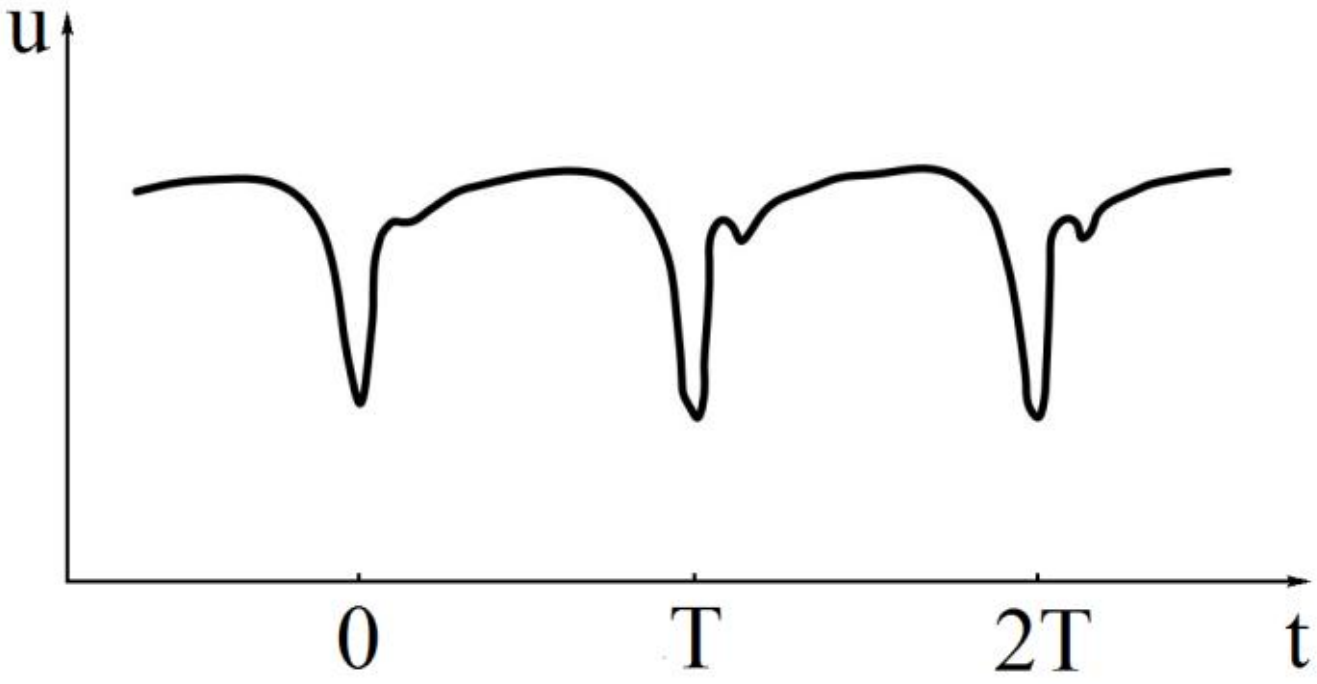

Figure 7. Experimental result of instantaneous streamwise velocity for the one-spike stage at a wall-normal position y/h=0.51 for plane Poiseuille flow (adapted from Nishioka et al. [21]). The y-axis is measured from the lower wall of the channel and h is the half-depth of the channel. The Reynolds number based on the half-width of channel and the centerline velocity of laminar flow is 5000. The symbol T expresses the period of the disturbance imposed on the incoming flow.

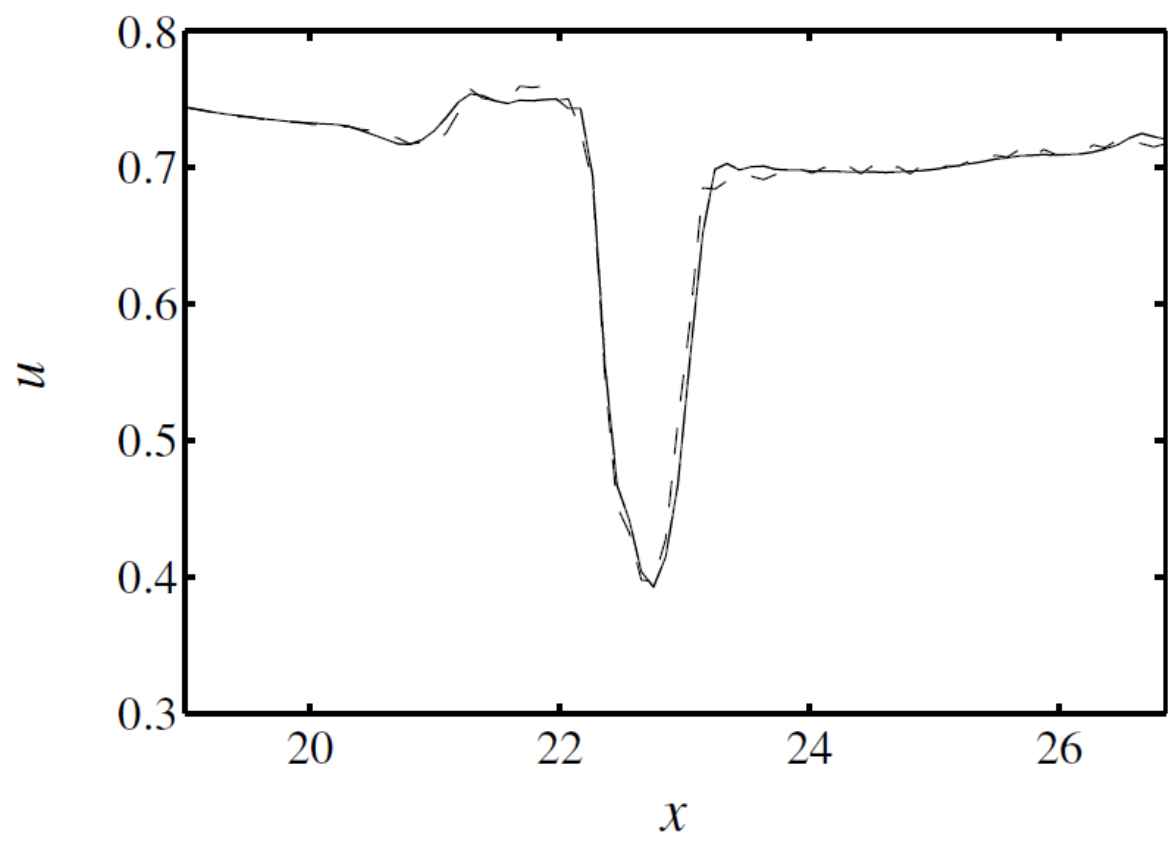

Figure 8. Instantaneous streamwise velocity for the one-spike stage at a wall-normal position *y/h*=-0.47 in the fluctuation peak plane about the spanwise direction for channel flow [24], with a grid resolution 1024 × 64 × 65, high-resolution ADM-RT, no-model LES (low-resolution spatial DNS). The Reynolds number based on the half-width of channel and the centerline velocity of laminar flow is 5000. This velocity profile is associated with the passage of the first hairpin vortex. At the walls, *y/h*=± 1, and *h* is the half-width of channel. Here, the ADM-RT format refers to a model based on the relaxation term (RT) of the approximate deconvolution model (ADM).

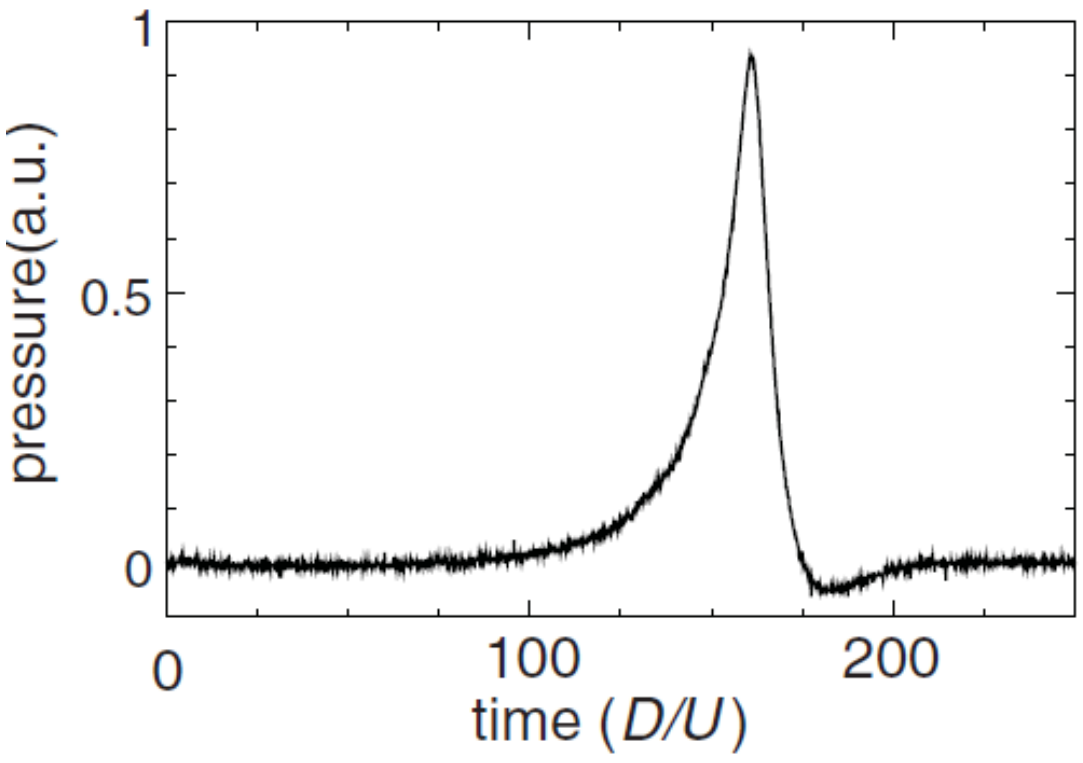

Figure 9. Measured pressure distribution in a puff of pipe flow experiment [44]. A puff of pipe flow is similar to a turbulent spot in plane Poiseuille flow or channel flow.

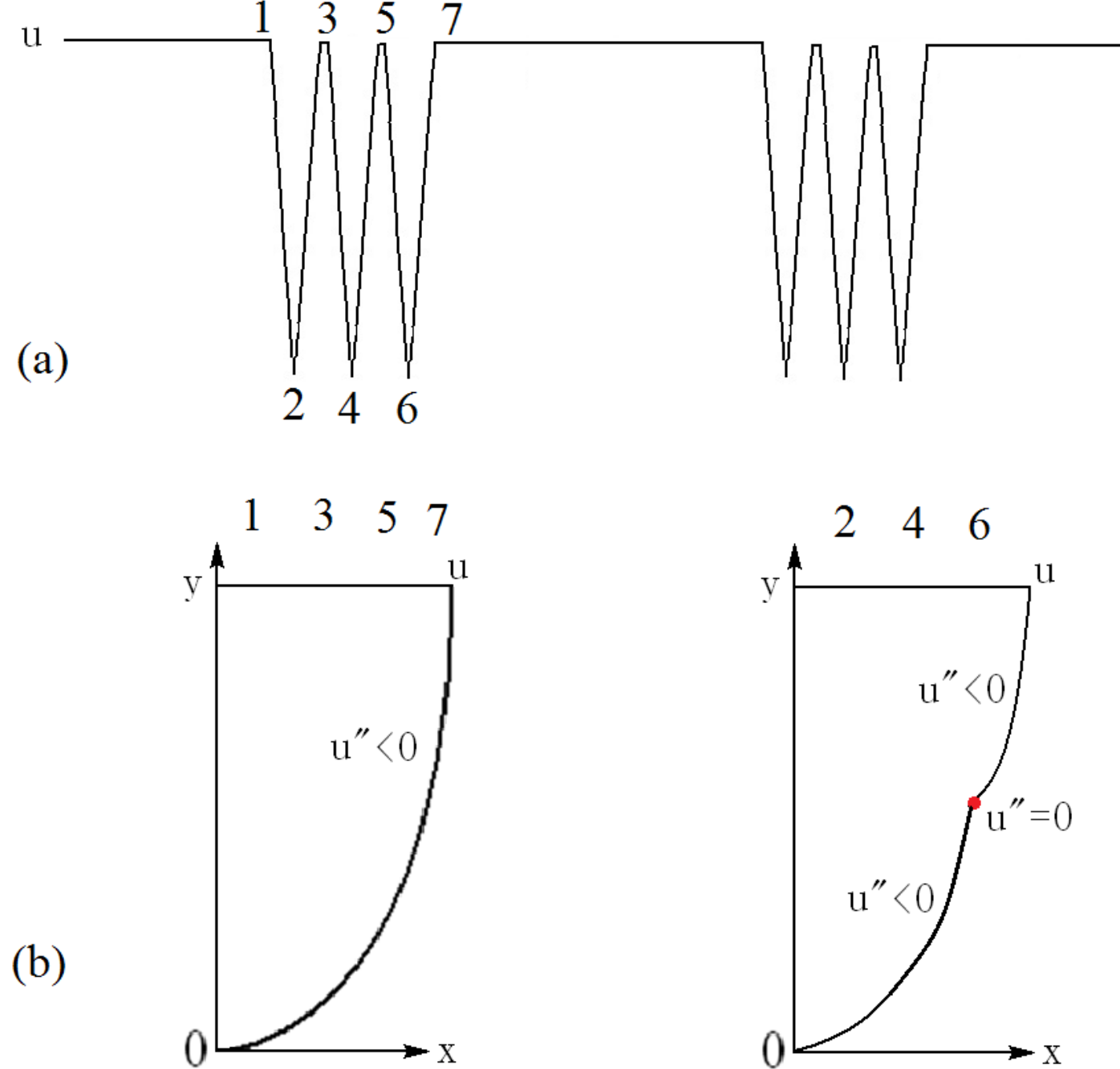

Figure 10. Explanation for the mechanism of generation of the stage of triple spikes. The oscillatory evolution of the shape of velocity profile along the streamwise direction leads to velocity discontinuities, which result in multiple velocity spikes. (a) Schematic of time tracing of velocity. (b) Schematic of velocity profile for the 7 points in (a).

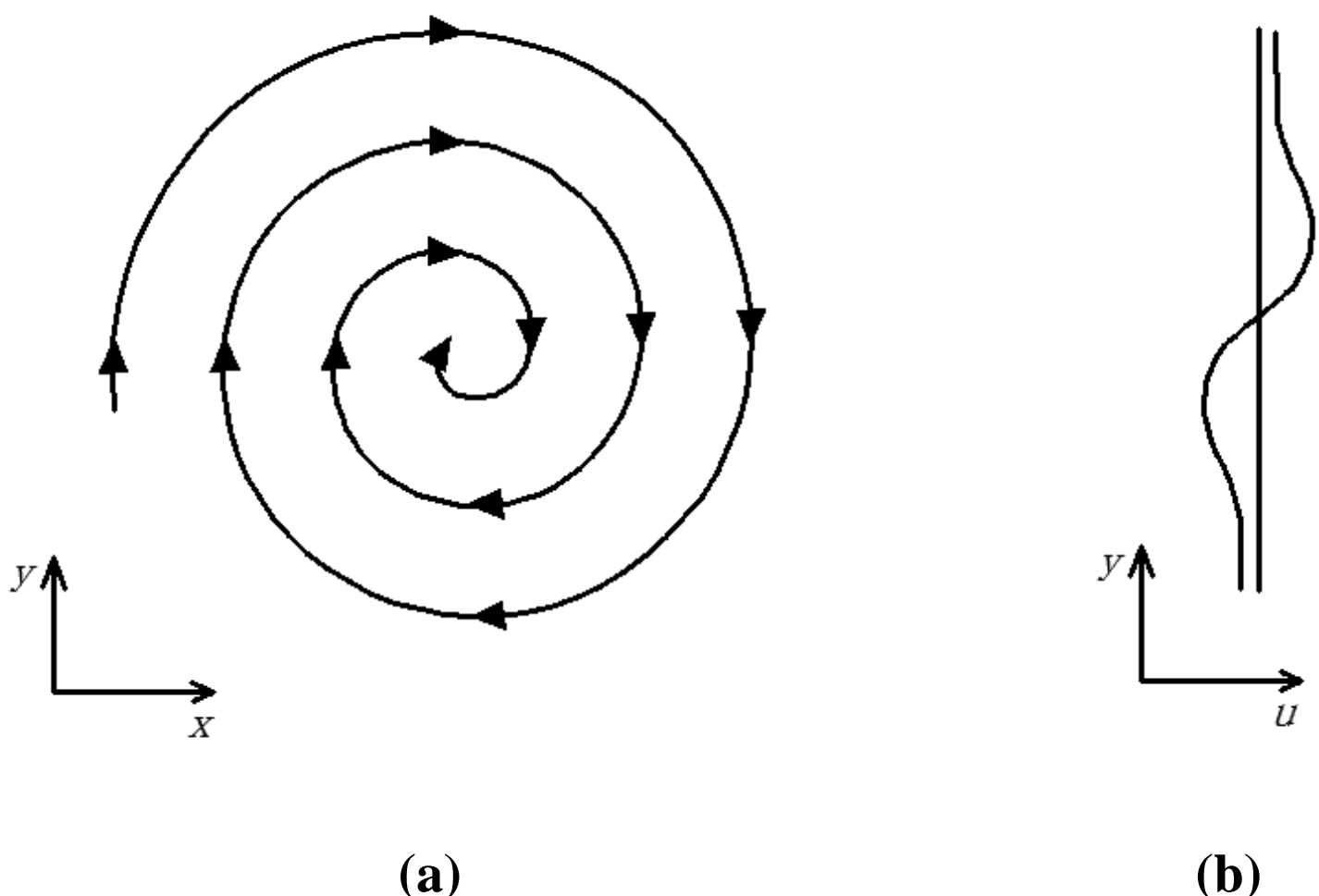

**(a)** **(b)**

Figure 11. Schematic of (a) a vortex and (b) the induced streamwise velocity.

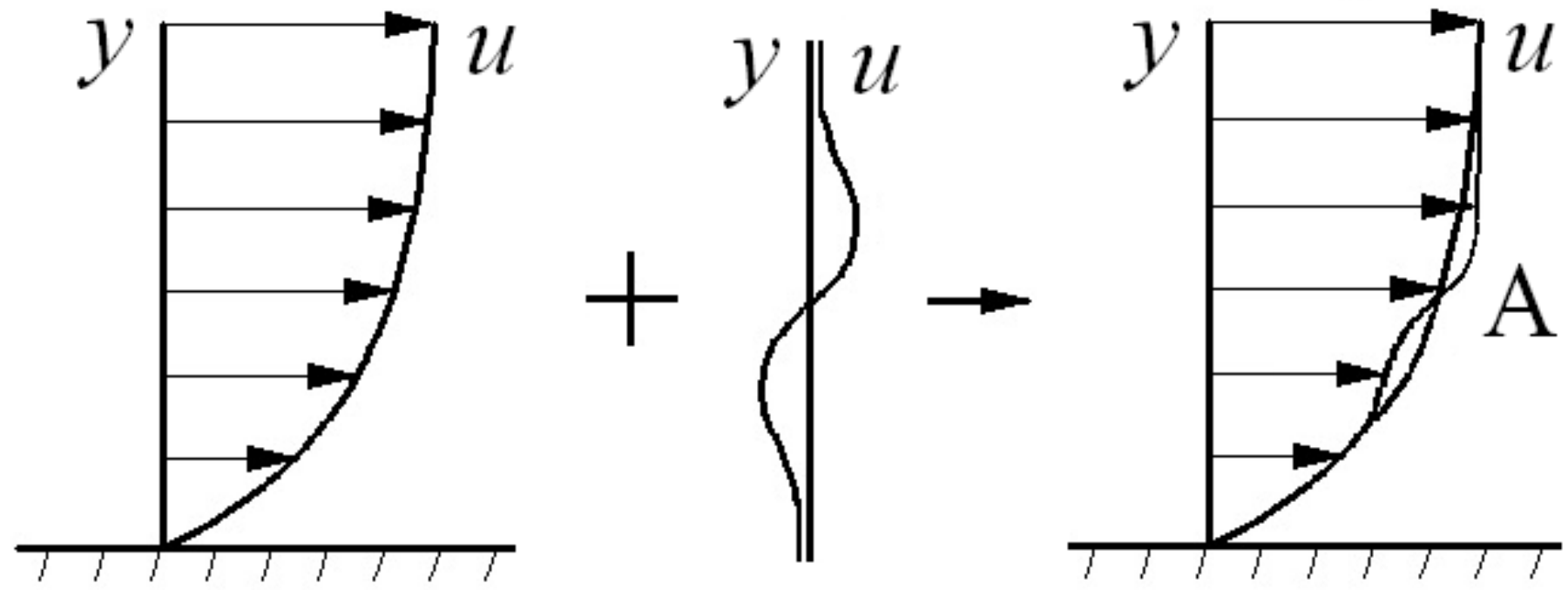

Figure 12. Streamwise velocity profile resulted from the overlap of a vortex with the incoming streamwise velocity. The role of vortices is to make the incoming velocity being inflectional.